\documentclass[letterpaper, 10 pt, conference]{ieeeconf}  

\IEEEoverridecommandlockouts                              
\usepackage[letterpaper, left=0.68in, right=0.68in, bottom=0.75in, top=0.75in]{geometry}
\usepackage{graphicx}

\title{\LARGE \bf
Learning Algorithms in Static Analysis of Web Applications
}

\author{Akash Nagaraj*, Bishesh Sinha, Mukund Sood, Yash Mathur, Sanchika Gupta, Dinkar Sitaram\\
Department of Computer Science, PES University\\
Bangalore, India\\
{\tt\small [akashn1897, mukundsood2013, mathuryash5]@gmail.com, [sanchikagupta, dinkars]@pes.edu}
}

\begin{document}

\maketitle
\thispagestyle{empty}
\pagestyle{empty}

\begin{abstract}
Web applications are distributed applications, they are programs that run on more than one computer and communicate through a network or server. This very distributed nature of web applications, combined with the scale and sheer complexity of modern software systems complicate manual security auditing, while also creating a huge attack surface of potential hackers. These factors are making automated analysis a necessity. Static Application Security Testing (SAST) is a method devised to automatically analyze application source code of large code bases without compiling it, and design conditions that are indicative of security vulnerabilities. However, the problem lies in the fact that the most widely used Static Application Security Testing Tools often yield unreliable results, owing to the \textit{false positive} classification of vulnerabilities grossly outnumbering the classification of \textit{true positive} vulnerabilities. \par 
This is one of the biggest hindrances to the proliferation of SAST testing, which leaves the user to review hundreds, if not thousands, of potential warnings, and classify them as either actionable or spurious. We try to minimize the problem of false positives by introducing a technique to filter the output of SAST tools. The aim of the project is to apply learning algorithms to the output by analyzing the true and false positives classified by OWASP Benchmark, and eliminate, or reduce the number of false positives presented to the user of the SAST Tool.
\end{abstract}

\textbf{\textit{ Keywords- machine learning; static application security testing; web applications.}}

\section{Introduction}

Owing to the widespread use of web applications, static analysis has emerged as a promising solution for automated security auditing. Web applications can scale to millions of lines of code which raises the chances of subtle security vulnerabilities occurring in the code. In particular, static analysis has shown great value when applied to information-flow vulnerabilities. These vulnerabilities also happened to be the most serious and prevalent forms of vulnerabilities in today’s landscape of web applications, which are divided into two main categories: integrity and confidentiality violations.\par
Security testing means to test for security vulnerabilities in software. By the increasing use of software and web based applications in more and more areas of our daily life, there is a direct correlation between the amount of sensitive data which is produced, and the amount of sensitive data that needs to be processed automatically, e.g., in electronic health or e-government applications. \par
Integrity violations include violations like cross-site scripting (XSS), where unvalidated or "unsanitized" input may execute on the victim's browser; cross-application scripting (XAS), where tags injected into widgets can alter their appearance and behavior; and SQL injection (SQLi), whereby unintended SQL commands are executed with malicious intent. Confidentiality violations, on the other hand are a dual problem. They arise when sensitive data emanating from an application is released to unauthorized observers. Security testing aims at checking applications for the presence of such violations. Due to the accessibility of modern service-oriented systems, security testing has gained much interest in the last few years, and has become a very promising field of research. \par
There are four possible test outcomes, when a tool is used to automatically detect a vulnerability in code:
\begin{itemize}
\item True Positive  - Real Vulnerability is correctly identified
\item False Negative - Real Vulnerability is wrongly ignored
\item True Negative  - False Alarm is correctly ignored
\item False Positive - False Alarm is wrongly identified
\end{itemize}

Although extensive research has been done in the field of static source code analysis, there in one major hurdle, that is the presence of a large number of false positives in the results generated by present-day static source code analyzers, which makes the predictions highly unreliable. The main aim of our project is to reduce the number of false positives generated by static source code analyzers.\par

Our approach is different from the previous attempts to solve the problem, using learning algorithms, as opposed to domain oriented predefined knowledge that only an expert could acquire. We have analyzed the source code of the web application in consideration, keeping key vulnerability signatures in mind. \par

We have implemented the elimination of false positives for the most common security as described by OWASP's list of vulnerabilities \cite{c1}:

\begin{enumerate}
\item Injection Attacks
	\begin{itemize}
    \item SQL Injection
    \item Command Injection
	\end{itemize}
\item Broken Authentication and Session Management
\item Sensitive Data Exposure
\item Broken Access Control
\item Cross-Site Scripting
\item Insecure De-serialization
\end{enumerate}

For this project, we worked with Java code deployed in web applications, as it is one of the most popular languages, with numerous web frameworks such as Spring, and JSF. This being said, it is important to note our approach can be extended to support any platform, and is not language dependent. To actually test the vulnerabilities generated by currently available static analysis tools, we performed a comparative study of all the currently available open-source vulnerability detection tools. We also used OWASP Benchmark, an open-source and open test suite designed to evaluate the speed, coverage, and accuracy of automated software vulnerability detection tools and services.\par

We used was 2,740 code snippets written in Java, from web applications and a static analysis tool was run with this data as the target. This resulted in the generation of an average of 3 vulnerability alerts per sample. We used this data along with the OWASP Benchmark~\cite{c2} analyzer's expected results, which was a list of \textit{True Positive} and \textit{False Positive} vulnerabilities found in the code. \par

The features we engineered for our learning model were  based on the embedded word vector representations of the actual line where the vulnerability was detected, and this was the input to our individual learning models, as well as the ensemble.

\section{Related work}
\subsection{A Classification for Model-Based Security Testing~\cite{c3}}

This paper first gives an overview of existing security testing approaches, and based on that, develops a classification for model-based security tests along the two dimensions risk and automated test generation. The classification allows for understanding which
areas of model-based security testing are already well-covered by research and practice, and furthermore, can serve as a guideline
for deciding which testing approach fits specific circumstances. Based on the classification, it identifies tasks for potential future research.\par
The takeaway from this paper is the various security approaches currently in use, as well as the benefits of model-based testing, which is a widely used method. It especially focuses on the integration of risks into their test models which has not been investigated in detail, but has high potential for practical application in security testing.

\subsection{ALETHEIA: Improving the Usability of Static Security Analysis~\cite{c4}}

The idea behind this paper is to apply statistical learning to the warnings output by the analysis based on user feedback on a small set of warnings, an idea very similar to ours. The approach used is a more interactive (and cumbersome) one, whereby the user classifies a small fragment of the issues reported by the analysis, and the learning algorithm then classifies the remaining warnings automatically. This is a user-centric approach which could be highly inaccurate, depending on the skill level of the user. \par
From this paper, we can learn one of the ways we could possibly target the problem in consideration. While the approach they have used gives the user a finer level of control over the output generated, not only is it a cumbersome solution, it could also be potentially inaccurate depending on the user, and is not a standard solution to the problem.

\subsection{Finding Security Vulnerabilities in Java Applications with Static Analysis~\cite{c5}}

This paper proposes a static analysis technique for detecting many recently discovered application vulnerabilities such as SQL injections, cross-site scripting, and HTTP splitting attacks.They have proposed a static analysis approach based on a scalable and precise technique. User-provided specifications of vulnerabilities are automatically translated into static analyzers. The approach claims to find all vulnerabilities matching a specification in the statically analyzed code. Results of our static analysis are presented to the user for assessment in an auditing interface integrated within Eclipse, a popular Java development environment.\par
One positive point about the approach used in this paper is that they have very few number of false positives. From this paper, we see yet another way to handle static source code testing. This paper is the closest to ours, as they are also dealing with web applications written in Java, but the downside is that it is not a generalizable technique, and is highly language dependent.

\section{Approach}

Security tools are amazing when they find a complex vulnerability in your code. But with widespread misunderstanding of the specific vulnerabilities that automated tools cover, end users are often left with a false sense of security. To determine how the tools really are at discovering and properly diagnosing security problems in applications, the test suite tests both real and fake vulnerabilities. The four possible outcomes of the benchmark are given above in the Introduction section.

Our first major challenge was generating the dataset. We evaluated many tools over the Benchmark, and finally settled on using the \textit{FindSecBugs} tool. The FindSecBugs tool would take each of the 2.740 files and perform static testing on them. If there was more than one vulnerability, it would provide information regarding all of them, in the order that it found the vulnerabilities. The output was provided in an XML file that followed a specific format. It gave us information regarding the name of vulnerability, the type of the vulnerability, the class in which the vulnerability appeared, the method in the class in which it appeared, and then all the lines that led to the code being vulnerable and a few others. Since the XML file had a specific format, we were able to build a parser and extract information from the fields that were relevant to us. The fields that we extracted from the file were:
\begin{enumerate}
\item Filename
\item Vulnerability Name
\item Type of Vulnerability
\item Line Number of Vulnerability
\end{enumerate}

Using this information, the parser would then open the appropriate file and extract the lines flagged by the tool as possible vulnerabilities. At this stage the fields that we extracted and place in our data set were all the fields extracted as mentioned above. We also added a column which had the category of the vulnerability based on the \textit{expected results} file part of the OWASP Benchmark. The categories are: 
\begin{itemize}
\item \textbf{pathtraver} - vulnerable path traversal
\item \textbf{hash} - hash value/message digest being insecurely passed
\item \textbf{trustbound} - trust bound vulnerability
\item \textbf{crypto} - weak implementation of a cryptographic algorithm used for encryption
\item \textbf{cmdi} - Command injection attack
\item \textbf{sqli} - SQL injection attack 
\item \textbf{weakrand} - use of a weak random number generator
\item \textbf{ldapi} - Lightweight Directory Access Protocol based vulnerabilities
\item \textbf{xss} - Cross Site Scripting vulnerabilities
\item \textbf{securecookie} - Cookie based vulnerabilities
\item \textbf{xpathi} - xpath injection
\end{itemize}

The final dataset that we used was a join between our previous dataset, and the \textit{expected results} file given by the OWASP Benchmark on the \textit{filename} and \textit{category} fields, with the classification as either a True positive or a False positive. \par
\begin{figure}[]
	\centering
  \includegraphics[width=\linewidth]{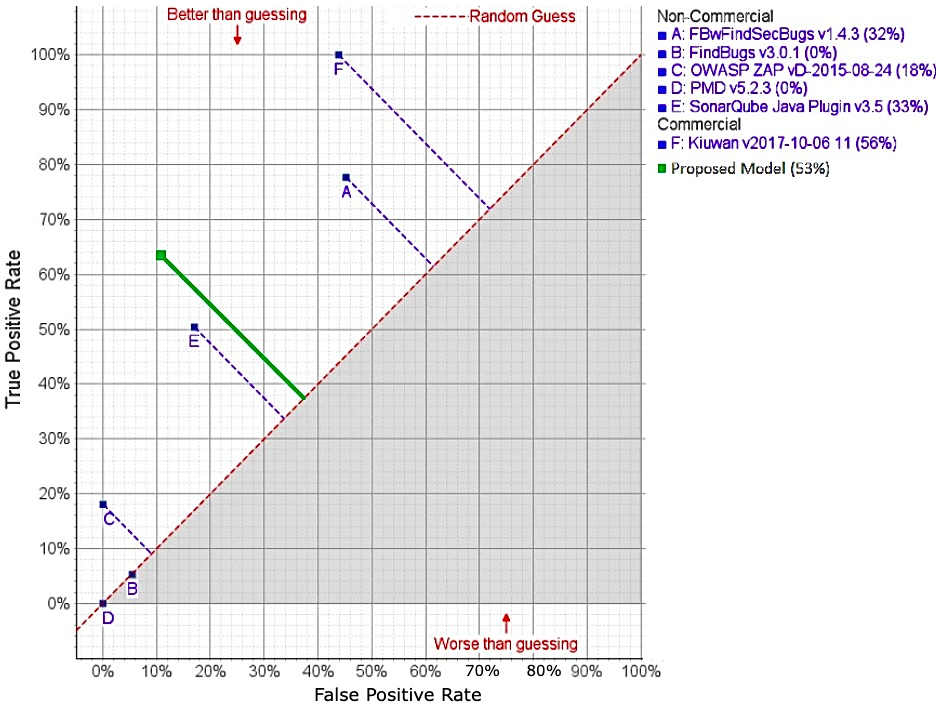}
  \caption{Comparison with pre-existing models}
\end{figure}
\begin{figure}[b!]
	\centering
  \includegraphics[width=6.0cm]{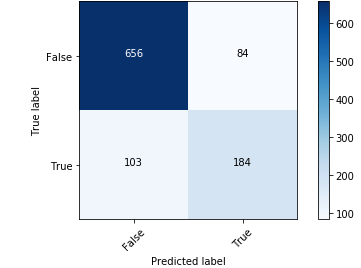}
  \caption{Confusion Matrix}
\end{figure}
\section{EXPERIMENTS AND RESULTS}
The next challenge was to build a classifier to distinguish between files which had a real vulnerability versus files that did not. To do this we thought of a few different ways, but finally settled on using a Word Embedding Model. We looked at two possible implementations, Word2Vec and pre-trained GloVe~\cite{c6} vectors. We quickly ruled out the possibility of using the pre-trained GloVe vectors, because these were trained on everyday conversation and text, and considering we were dealing with code blocks, it would make no sense to use them, because a relationship between the vector representation of these words would have no relation to those same words used in a programming language context. We hence used the Word2Vec Model trained on our own corpus of code blocks extracted from the Benchmark test cases. \par
 
Once we trained the Word2Vec Model, we used vector averaging. For each code block, we loaded the vectors for each word and averaged them across the number of words in that particular code block to give us a single vector that would represent the entire code block.  \par

The Machine Learning Models that we built were a Support Vector Machine [7] and a Random Forest Classifies, and finally we also tried out XGBoost and created an ensemble out of these three Models. \par

Table I summarizes the results achieved, with varying vector dimensions for all the different models we have used.

\begin{table}
\caption{Summary of Results}
\label{Results}
\begin{center}
\begin{tabular}{@{}llllll@{}}
\hline
\textbf{Model}      &\textbf{Dim 30}		&\textbf{Dim 20}	&\textbf{Dim 10} 					\\
Random Forests 		& 81.89					& 82.38				& 82.77    							\\
SVM 				& 73.61   				& 74.00  			& 75.82								\\	
XGBoost				& 81.50   				& 81.40				& 82.38       						\\
                    &    	  				&        			&  									\\
Ensemble    		& 81.79 				& 82.18				& 81.57 							\\ \hline
\end{tabular}
\end{center}
\end{table}

\section{CONCLUSIONS}

As shown in Table I, our model accurately reduces the number of false positive warnings generated, and brings it down from 4,970 to 447, which reduces the number of false positives from 73\% to a mere 6.5\%, which is far better when compared to the other pre-existing techniques currently in use.

\section{Reproducible Research}
In the spirit of reproducible research, the work done is publicly available. The Python code for all the experiments and work done in this paper can be accessed at~\cite{c8}.

\addtolength{\textheight}{-12cm}   


\end{document}